\documentclass[aps,preprint,showpacs,amsmath,amssymb]{revtex4}
\usepackage{epsfig}
\usepackage{wasysym}
\usepackage{amssymb}

\begin{document}

\newcommand{\bvec}[1]{\mbox{\boldmath ${#1}$}}
\title{Coupling strength of the $N^*(1535)S_{11}$ to the $K^+\Lambda$ channel}
\author{T. Mart}
\affiliation{Departemen Fisika, FMIPA, Universitas Indonesia, Depok 16424, 
  Indonesia}
\date{\today}
\begin{abstract}
  The $N^*(1535)S_{11}$ coupling strength to the $K^+\Lambda$ channel,
  $g_{N^*(1535)\Lambda K^+}$, 
  is extracted from  the latest and largest $K^+\Lambda$ photoproduction
  database by using an isobar model. It is found that the coupling is small.
  In term of the coupling ratio the best result is 
  $R\equiv |g_{N^*(1535)\Lambda K^+}/g_{N^*(1535)\eta p}|
  =0.460\pm 0.172$, much smaller than that obtained from the
  isobar analysis of the $J/\psi$ decays, i.e., $1.3\pm 0.3$, 
  but consistent with the results of the unitary chiral 
  approach of the same decay processes as well as the partial 
  wave analysis of kaon photoproduction, 
  i.e., $R=0.5\sim 0.7$. The different results of $R$  found here and in
  literature to date suggest that a more solid definition of the
  coupling constant, especially in the case of $N^*(1535)S_{11}$ state,
  is urgently required, before a fair comparison can be made.
\end{abstract}

\maketitle

The nucleon resonance $N^*(1535)S_{11}$ occupies a special place
in the Particle Data Group (PDG) listing, because it has an 
extraordinarily large branching fraction to the $\eta N$ channel,
i.e., 45$-$60\%  \cite{pdg}. To explain this a number of 
mechanism has been
proposed, e.g., the $K\Lambda$-$K\Sigma$ 
\cite{kaiser} and meson-baryon \cite{inoue-2002}
quasi-bound states, as well as the introduction 
of pentaquark component in the nucleon
besides the conventional $uud$ state \cite{zou_npa}. 
At the quark level,
the existence of the $N^*(1535)S_{11}$ leads to a long standing 
problem, since the conventional constituent quark model 
predicts this resonance to be the lowest mass state
\cite{capstick98}, in contrast
to the fact that there exists an $N^*(1440)P_{11}$ state
with $J^p=1/2^+$ as listed in the Review
of Particle Physics by PDG \cite{pdg}. A suggested
solution to this mass reverse problem is the introduction of 
the pentaquark admixture \cite{zou_npa}. However, such 
mechanism leads to a large $N^*(1535)S_{11}$  coupling
to the $K\Lambda$ channel.
More dramatically, by analyzing the $J/\psi\to{\bar p}K^+\Lambda$ 
and $J/\psi\to{\bar p}p\eta$ experimental data within an isobar
model it was found that the coupling of the $N^*(1535)S_{11}$ to the
$K\Lambda$ channel is larger than its coupling to the $\eta N$
channel, i.e., 
$R\equiv |g_{N^*(1535)\Lambda K^+}/g_{N^*(1535)\eta p}|=1.3\pm 0.3$ 
\cite{Liu:2005pm}. 
A direct consequence of this large ratio is
that the mass and width of the $N^*(1535)S_{11}$ should be
1400 and 270 MeV, respectively. Clearly, this is an 
unexpected result, since this finding is considered \cite{sibirtsev} to 
differ radically from the standard value \cite{pdg}.
However, we observe that in literature the values of this 
ratio are 
wildly scattered. For instance, the unitary chiral approach 
found $R=0.5\sim 0.7$ \cite{geng}. A similar result was obtained
by the partial wave analysis of kaon photoproduction \cite{Sarantsev:2005tg}, 
whereas a coupled-channels calculation
predicted $R=0.8\sim 2.6$ \cite{geng,penner}. On the other hand,
the result of the $s$-wave pion-nucleon scattering analysis 
in a unitarized chiral effective Lagrangian indicates that 
$|g_{N^*(1535)\Lambda K^+}|^2 > |g_{N^*(1535)\eta p}|^2$ \cite{Bruns:2010sv}. 
It is important to note that these different results originate 
from different methods of analysis as well as different data
and, as a consequence, there
exist conceptual differences between the extracted coupling
constant. Therefore, care must 
be taken when one wants to compare these results in term of a unique 
definition of coupling constant. 

With the accumulating precise 
kaon photoproduction data from the modern continues
electron beam facilities such as CEBAF, ELSA, SPring-8, 
ESRF, and MAMI, it is 
obviously important to consider the $\gamma p\to K^+\Lambda$
process for extracting the $g_{N^*(1535)\Lambda K^+}$ coupling, since the
$g_{N^*(1535)\eta p}$ coupling is relatively well known \cite{pdg}.
The process has been studied for decades by using a number of
phenomenological models. However, the isobar model is the most relevant
one for the present discussion. We note that most of the models did not
include the $N^*(1535)S_{11}$ resonance because its mass is located
below the reaction threshold. Other models include this resonance
mainly because kaon photoproduction is one of 
the coupled channels being analyzed. Interestingly, 
the conclusions from these kaon photoproduction studies vary 
from one analysis to another. 
For instance, Refs. \cite{maxwell,chiang} found that the resonance
is  less important in the $K^+\Lambda$ photoproduction, 
whereas Refs. \cite{Julia-Diaz:2006is,shklyar} drew 
an opposite conclusion.

Given the critical consequence of large $R$ 
in many aspects of hadronic physics, we believe that
it is urgent to extract the value 
from kaon photoproduction data. In this paper we report on 
the result of 
this extraction, which makes use of an isobar model 
constructed from appropriate Feynman diagrams based on
our previous but latest model \cite{mart2012}. 
The model consists of the standard \mbox{$s$-}, $u$-, and $t$-channel Born terms 
along with the $K^{*+}(892)$, $K_1(1270)$ vector 
mesons and the $\Lambda^*(1800)S_{01}$, $\Lambda^*(1810)P_{01}$ hyperon
resonance. In the $s$-channel the model takes the $N^*(1650)S_{11}$,  
$N^*(1700)D_{13}$, $N^*(1710)P_{11}$, $N^*(1720)P_{13}$, $N^*(1840)P_{11}$, 
$N^*(1900)P_{13}$, 
$N^*(2080)D_{13}$, $N^*(2090)S_{11}$, and  $N^*(2100)P_{11}$ nucleon resonances
into account. Note that the choice of these nucleon resonances
is consistent with the result of the partial wave analysis
\cite{Sarantsev:2005tg} and the
2012 PDG listing \cite{pdg}. We have also compared our resonance
configuration with that used by the Ghent group \cite{decruz}
and found that our configuration is consistent up to spin 5/2
resonances. The omission of spin 5/2 resonances in our model
was discussed in Ref.~\cite{mart2012}.

To approximately account for unitarity corrections 
at tree-level we use energy-dependent widths along with partial 
branching fractions in the resonance propagators 
\cite{frank}. Furthermore, to account for the fact that 
hadrons are composite objects, hadronic form factors are considered
in hadronic vertices, where the gauge invariance of the amplitude
after the form factor inclusion is restored by using
the Haberzettl prescription \cite{Haberzettl:1998eq}.
The model fits all latest $K^+\Lambda$ photoproduction data
consisting of differential cross section 
\cite{Bradford:2005pt,mcCracken,Sumihama:2005er,Hicks_2007}, 
recoil polarization \cite{Bradford:2005pt,lleres07},
beam-recoil double polarization \cite{Bradford:2007,lleres09}
as well as photon $\Sigma$ and target $T$ asymmetries \cite{lleres07}
data. In total, our database consists of more than 
3500 data points. To our knowledge, this is the largest $K^+\Lambda$
photoproduction database intended for the present purpose.
Two different models (A and B) were proposed in Ref.~\cite{mart2012}.
Both models use the same resonance configuration as described
above, but in the model A [B] the mass and width of 
the $N^*(2080)D_{13}$ [$N^*(1900)P_{13}$] resonance 
were considered as free parameters. As a result a total $\chi^2$ of
9084 (9494) was obtained in model A (B). For a more detailed discussion
on the performance of both models we refer the reader to 
Ref.~\cite{mart2012}.

\begin{figure}[t]
  \begin{center}
    \leavevmode
    \epsfig{figure=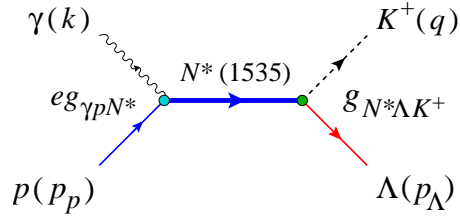,width=60mm}
    \caption{(Color online) Contribution of the $N^*(1535)S_{11}$ 
      diagram to the kaon photoproduction process.
      }
   \label{fig:s-channel} 
  \end{center}
\end{figure}

In the present work we include the $N^*(1535)S_{11}$ contribution in
both models and refit the experimental data to the model calculations
by adjusting all coupling strengths. 
The Lagrangian for the magnetic transition of this resonance reads
\cite{feuster}
\begin{eqnarray}
  \label{eq:magnetic}
  {\cal L}_{\gamma NN^*}=\frac{e}{4m_N}{\bar \psi}_{N^*}g_{\gamma NN^*}
  \gamma_5\sigma_{\mu\nu}\,\psi_N\, F^{\mu\nu}+{\rm H.c.} ,
\end{eqnarray}
where $F^{\mu\nu}=\partial^\mu A^\nu-\partial^\nu A^\mu$ and
other terms are self explaining. In 
the $K\Lambda N^*$ vertex one obtains
\cite{Liu:2005pm} 
\begin{eqnarray}
  \label{eq:hadronic-Lagrangian}
  {\cal L}_{N^*\Lambda K}=-ig_{N^*\Lambda K} {\bar \psi}_\Lambda
  \Phi_K\,\psi_{N^*} + {\rm H.c.} .
\end{eqnarray}
The corresponding Feynman diagram is depicted in 
Fig.\,\ref{fig:s-channel}. 
Obviously, only the product of 
$g_{\gamma pN^*}\, g_{N^*\Lambda K}$ can be extracted from the
fitting process. Nevertheless, the $g_{\gamma pN^*}$ coupling
is relatively well known from the PDG value of the helicity 
amplitude. 
For a negative-parity spin 1/2 resonance this 
amplitude can be related to the 
electromagnetic coupling constant via \cite{feuster}
\begin{eqnarray}
 A_{1/2}^p &=& \frac{1}{2m_p}
         \left(\frac{m_{N^*}^2-m_p^2}{2m_p}\right)^{1/2}~ eg_{\gamma pN^*} ~.
\end{eqnarray}
Using $A_{1/2}^p=0.090\pm 0.030$ GeV$^{-1/2}$ \cite{pdg} 
we obtain 
$g_{\gamma pN^*}=0.335 \pm 0.112$.

The photoproduction amplitude ${\cal M}$ is conventionally 
decomposed 
into the gauge- and Lorentz-invariant 
matrices $M_i$, 
\begin{eqnarray}
  \label{eq:decompose}
  {\cal M} = \sum_{i=1}^4 A_i(s,t,u)\, M_i ~,
\end{eqnarray}
where $s,t$, and $u$ are Mandelstam variables, 
since all observables can be calculated from $A_i$.
The explicit forms of $A_i$ and $M_i$ can be found, e.g.,
in Ref.~\cite{frank}. Note that, in the present paper 
models A1 and B1 refer to the two original models 
A and B explained above \cite{mart2012}
but after including the $N^*(1535)S_{11}$ resonance
in the model and refitting the experimental data. Recently, 
we found that by adding 
a $\Lambda^*(1600)P_{01}$ hyperon resonance in the 
original model A significant 
improvement can be achieved, i.e., 
the $\chi^2$ is greatly reduced and the background 
form factor cut-off is increased \cite{cepi}. 
In the present study we find that 
including the $\Lambda^*(1600)P_{01}$ in the original model A 
reduces the $\chi^2$ from 9084 to 8817. By
adding the  $N^*(1535)S_{11}$ to this result we obtain model A2
which has $\chi^2=8716$ as listed in  Table \ref{tab:coupling}.
As seen in this Table, the inclusion of the $\Lambda^*(1600)P_{01}$ 
simultaneously increases the 
Born cut-off, whereas the resonance one seems to be unaffected.

The problem of the over-damped Born terms due to the extremely
soft (small) Born cut off has been extensively discussed in literature 
(see e.g. Ref.~\cite{Bydzovsky:2006wy}). It was suspected that such a 
cut-off is artificial and the corresponding model is, therefore, 
far from a realistic
description of the process \cite{missing-d13}. 
This problem seems to appear in model B1,
for which the value of $\Lambda_{\rm Born}$ is less than one half
of that obtained in model A2. 

\begin{table}[t]
  \centering
  \caption{The $N^*(1535)S_{11}$ coupling constant $g_{N^*\Lambda K^+}$, the hadronic
    form factor cut-off $\Lambda$, and the coupling constant ratio $R$
    obtained from three different
    models in the present work. The product of electromagnetic and 
     hadronic couplings $g_{\gamma pN^*}\, g_{N^*\Lambda K^+}$ is
    obtained from refitting the experimental data.
    Also shown in this Table are the number of
    free parameters ($N_{\rm par.}$), the $\chi^2$, and the $\chi^2$ per
    number of degrees of freedom ($\chi^2/N_{\rm dof}$) for each model.}
  \label{tab:coupling}
  \begin{ruledtabular}
  \begin{tabular}[c]{lccc}
    Parameter & A1 & A2 & B1 \\
    \hline
    $g_{\gamma pN^*}\, g_{N^*\Lambda K^+}$ & $0.220\pm 0.033$ & 
    $0.286\pm 0.030$ & $0.157\pm 0.024$ \\
    $g_{N^*\Lambda K^+}$ & $0.656\pm 0.240$ & $0.853\pm 0.298$ &
    $0.469\pm 0.172$\\
    $R$ & $0.354\pm 0.138$ & $0.460\pm 0.172$ & $0.253\pm 0.099$ \\
    $\Lambda_{\rm Born}$ (GeV) & 0.920 & 1.070 & 0.483 \\
    $\Lambda_{\rm Res.}$ (GeV) & 1.356 & 1.364 & 1.460 \\
    [1ex]
    \hline
    $N_{\rm par.}$ & 30 & 31 & 30 \\
    $\chi^2$          & 9035 & 8716 & 9480 \\
    $\chi^2/N_{\rm dof}$          & 2.555  & 2.466 & 2.681 \\
  \end{tabular}
  \end{ruledtabular}
\end{table}

We note that there is a strong correlation between the
$N^*(1535)S_{11}$ and $N^*(1650)S_{11}$ resonances. This
is indicated by the large value of the correlation coefficient
of these resonances given by the Minuit output  
(e.g., about 0.94 for model A2). Nevertheless, we found 
that the extracted couplings of both resonances
are comparably small.

\begin{figure}[t]
  \begin{center}
    \leavevmode
    \epsfig{figure=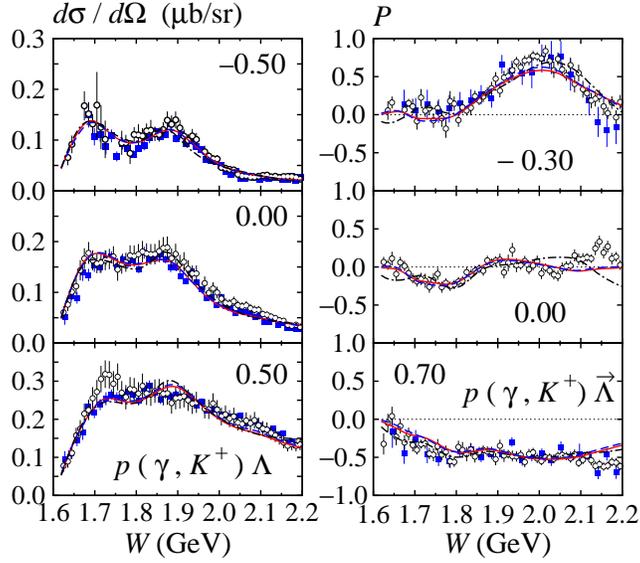,width=85mm}
    \caption{(Color online) Differential cross sections 
      $d\sigma/d\Omega$
      and recoil polarization observables $P$ 
      sampled at three different kaon angles $\theta$ 
      for the three models given in
      Table~\ref{tab:coupling}. The value of $\cos\theta$ is given
      in each panel. The solid, dashed, and dash-dotted lines 
      correspond to the models A1, A2, and B1, respectively.
      Experimental data are from the 
      CLAS collaboration (solid squares \cite{Bradford:2005pt}
      and open circles \cite{mcCracken}).}
   \label{fig:dkpl_pol} 
  \end{center}
\end{figure}

\begin{figure*}[t]
  \begin{center}
    \leavevmode
    \epsfig{figure=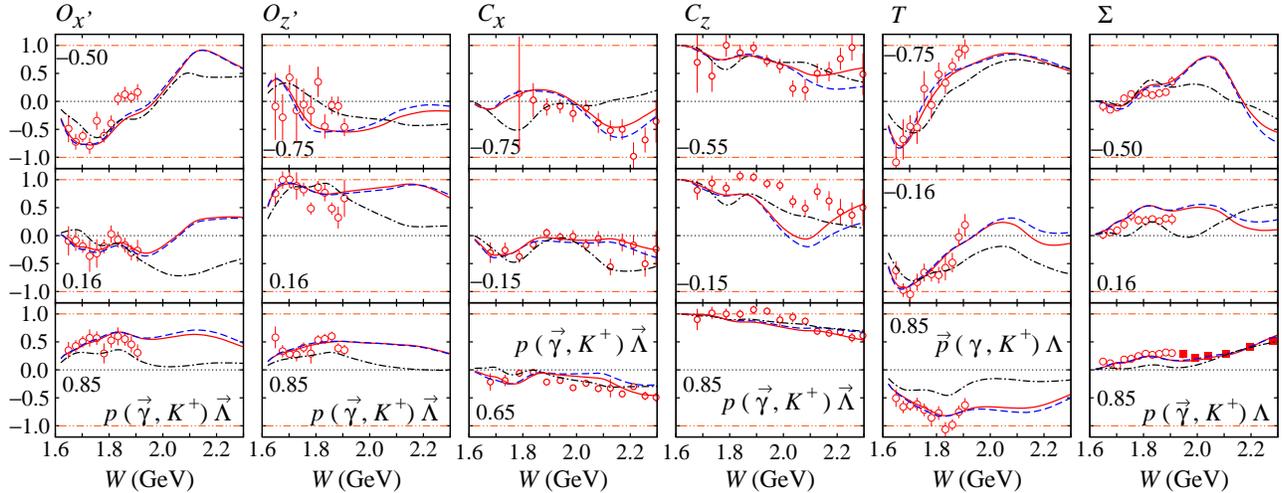,width=170mm}
    \caption{(Color online) Same as Fig.~\ref{fig:dkpl_pol}, but for the
      beam-recoil double polarization observables $O_{x'}$, $O_{z'}$,
      $C_x$, and $C_z$, 
      as well as the target $T$ and beam $\Sigma$ asymmetries. 
      The notation of the curves is as in Fig.~\ref{fig:dkpl_pol}.
      Experimental data are from the 
      GRAAL \cite{lleres09,lleres07},
      CLAS \cite{Bradford:2007}, and LEPS collaborations
      \cite{Sumihama:2005er}.}
   \label{fig:ox_oz_cx_cz_tar} 
  \end{center}
\end{figure*}

The performance of the presented models in describing
the selected experimental data is displayed in 
Figs.~\ref{fig:dkpl_pol} and \ref{fig:ox_oz_cx_cz_tar}.
Obviously, models A1 and A2 are superior to 
model B1. Although the difference between models A1 and A2
in the differential cross section
and recoil polarization data is graphically 
subtle, we find that numerically the 
presence of the $\Lambda^*(1600)P_{01}$ in model A2 
improves the agreement with data. Sizable improvements are also
found in the double polarization observables as well as
in the target and photon asymmetries.
Furthermore, from Table \ref{tab:coupling} 
we can conclude that model A2 is the most reliable model
for our present purpose.

\begin{figure}[t]
  \begin{center}
    \leavevmode
    \epsfig{figure=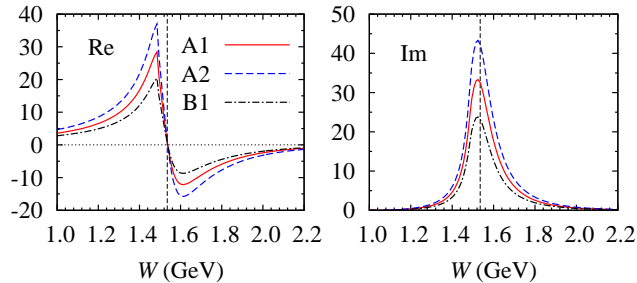,width=85mm}
    \caption{(Color online) Real and imaginary parts of the
      $A_1$ amplitude (in $10^{-3}$ GeV$^{-2}$) obtained from three different models.}
   \label{fig:a13-ir} 
  \end{center}
\end{figure}

The presence of the $N^*(1535)S_{11}$ in model A1, A2, and B1
improves the $\chi^2$ from 9084 to 9035, from 8817 to 8716,
and from 9494 to 9480, respectively. Therefore, the 
presence of the $N^*(1535)S_{11}$ 
has the strongest effect in model A2. This effect is closely
related to the $N^*(1535)S_{11}$ coupling strength given in 
Table \ref{tab:coupling}. Since the contribution of 
this resonance to the $A_i$ in Eq.~(\ref{eq:decompose}) 
is $A_2=0$, and $A_3=A4\propto A_1$, information on the 
$A_1$ displayed in Fig.~\ref{fig:a13-ir} 
is sufficient for estimating the $N^*(1535)S_{11}$ effect. 
The different magnitudes of $A_1$ exhibited by the three models
in Fig.~\ref{fig:a13-ir} originates 
from the different $g_{N^*\Lambda K^+}$ magnitudes
shown in Table \ref{tab:coupling}. 
Thus the larger the $g_{N^*\Lambda K^+}$ value, the more
important the contribution of this resonance to the photoproduction
process. 

\begin{figure}[b]
  \begin{center}
    \leavevmode
    \epsfig{figure=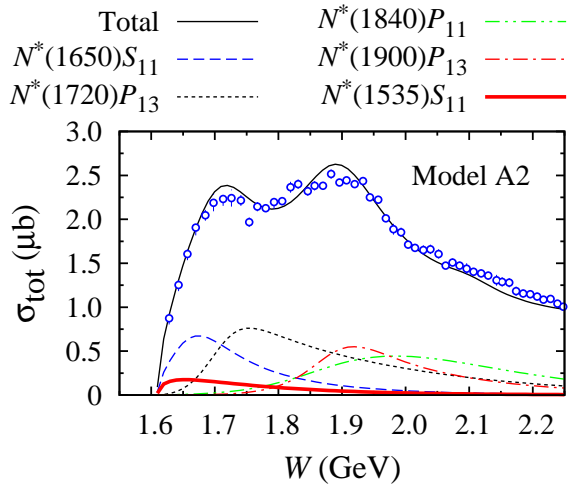,width=80mm}
    \caption{(Color online) Contributions of the most important
      resonances to the total cross section 
      of the $\gamma p\to K^+\Lambda$ process for Model A2.}
   \label{fig:contrib} 
  \end{center}
\end{figure}

Although the strongest effect of the $N^*(1535)S_{11}$ 
appears in model A2,
comparison of the $N^*(1535)S_{11}$ contribution with those from 
other dominant resonances in this model is extremely important, especially 
when we consider the claim that the $N^*(1535)S_{11}$ dominates the
$pp\to pK^+\Lambda$ reaction \cite{Liu:2005pm}, whereas
the experimental Dalitz plot strongly
suggests the $N^*(1650)S_{11}$ as a dominant resonance at low
energies \cite{el-samad}. The comparison is given in
Fig.~\ref{fig:contrib}.
Obviously, the dominant contributors
near threshold are the $N^*(1650)S_{11}$ and $N^*(1720)P_{13}$
resonances. Compared with these two resonances the $N^*(1535)S_{11}$ 
contribution is significantly smaller.
This result corroborates the finding of previous
studies using isobar model \cite{maxwell,mart2}
and is in agreement with the result of the
Dalitz plot \cite{el-samad}. 

To obtain the ratio $R$ given in Table \ref{tab:coupling}
we need to calculate the 
$g_{N^*(1535)\eta p}$ coupling constant. Using the
branching ratio given by PDG, 
i.e., $\Gamma(N\eta)/\Gamma_{\rm total}=0.42\pm 0.10$
\cite{pdg} and the relation between the branching ratio
and the corresponding coupling constant 
\cite{feuster} 
\begin{eqnarray}
  \label{eq:decay}
  \Gamma_{N^*} = \frac{g_{N^*\eta N}^2}{4\pi}\, p\, \frac{E_N+ m_N}{\sqrt{s}} ,
\end{eqnarray}
where 
  $p=[\{s-(m_N+m_\eta)^2\}\{s-(m_N-m_\eta)^2\}]^{1/2}/{2s^{1/2}}$
and $E_N=(p^2+m_N^2)^{1/2}$, we obtain
  $| g_{N^*(1535)\eta p}| =1.853\pm 0.244$, 
which is in good agreement with the result obtained
in Ref.~\cite{geng}. Using this value and the extracted $g_{N^*\Lambda K^+}$
from kaon photoproduction 
we obtain the ratio $R$ for all three models and list them 
in Table \ref{tab:coupling}.  
Since model A2 has been shown to be the most reliable model,
we believe that the best result of the present work yields 
\begin{eqnarray}
  \label{eq:ratio}
  R=0.460\pm 0.172.
\end{eqnarray}
Note that the large error bar is mainly due to the large uncertainties
in the PDG values of the $A_{1/2}$ and 
$\Gamma(N\eta)/\Gamma_{\rm total}$ \cite{pdg}
(see also Table \ref{tab:coupling} for the error given
by the fit to kaon photoproduction data).
This result is obviously smaller than
that obtained in Ref.~\cite{Liu:2005pm}, i.e., $1.3\pm 0.3$, and slightly
below the lower bound of the coupled-channels result, i.e.
$0.8\sim 2.6$ \cite{geng,penner}. However, the present result is
consistent with the results from the unitary chiral approach of the
$J/\psi$ decays \cite{geng}
and the partial wave analysis of kaon photoproduction, 
i.e., $R=0.5\sim 0.7$. We also note that the product of 
$g_{\gamma pN^*} g_{N^*\Lambda K^+}$ extracted from the 
$K^+\Lambda$ photoproduction in a recent isobar analysis
\cite{maxwell} is even smaller than the 
present result (see Table II of Ref.~\cite{maxwell}). Very likely the
smaller value in Ref.~\cite{maxwell}
is due to the complication of various interfering
resonances (about 23 nucleon and hyperon resonances) in the model.
To understand the different values obtained from the unitary chiral 
approach \cite{geng} and the isobar model \cite{Liu:2005pm}, 
the former was interpreted as a measure of the strength 
of the $K\Lambda$ component in the $N^*(1535)S_{11}$ wave function, whereas
the latter was considered as the fitted coupling strength 
near the $K\Lambda$ threshold \cite{geng}.
The result of the present isobar model given in Eq.~(\ref{eq:ratio})
supports the small value of the $g_{N^*(1535)\Lambda K^+}$.
Nevertheless, as stated above, a more conclusive result 
should wait for a more solid
definition of the resonance coupling constant, since 
there exist conceptual differences in 
current and available extracted coupling constants. 
We note, however, that one of the possible solutions 
to this problem is the introduction of the residue of
the transition amplitude as explained in the Review 
of the $N$ and $\Delta$ Resonances of the 2012 Review
of Particle Physics \cite{pdg}. In principle, this
residue can be calculated from a contour integral of
the transition amplitude around the pole position
in the complex energy plane and the result is proportional
to the coupling constant. Thus, future extractions of 
the hadronic coupling constants should consider this
issue in a comprehensive way, i.e.,
it applies not only to the $g_{N^*(1535)\Lambda K^+}$
coupling, but also to both  $g_{\gamma pN^*}$ and
$g_{N^*(1535)\eta p}$ couplings used to derive $R$.

In conclusion, we have extracted the $g_{N^*(1535)\Lambda K^+}$
coupling strength and its ratio to the $g_{N^*(1535)\eta p}$
coupling from a large $K^+\Lambda$ photoproduction database
by means of an isobar model. The result is 
significantly smaller than that obtained from the isobar analysis
of $J/\psi$ decays, but comparable to the results of the 
unitary chiral calculation as well as the partial wave analysis.
We have indicated that the result is not conclusive due
to the inherent problem in the method.

The author thanks Bing Song Zou for encouraging this study.
This work has been supported in part by the University of Indonesia
and by the Competence Grant of the Indonesian 
Ministry of Education and Culture.

\end{document}